\documentclass{desyproc}

\usepackage{wrapfig}

\def\Journal#1#2#3#4{{#1} {\bf #2}, #3 (#4)}

\def\NIMA{{\em Nucl. Instrum. Methods} A}

\def\PLB{{\em Phys. Lett.} B}
\def\PRL{\em Phys. Rev. Lett.}
\def\PRD{{\em Phys. Rev.} D}

\def\IJMPA{{\em Int. J. Mod. Phys.} A}
\def\EPJC{{\em Eur. Phys. J.} C}

\def\NIMA{{\em Nucl. Instrum. Methods Phys. Res.} A}

\mathchardef\mhyphen="2D

\begin{document}
\title{Search for low-mass Dark Matter with the CRESST Experiment}

\author{{\slshape Holger Kluck\textsuperscript{4,5},
G.~Angloher\textsuperscript{1}, P.~Bauer\textsuperscript{1}, A.~Bento\textsuperscript{1,8}, C.~Bucci\textsuperscript{2}, L.~Canonica\textsuperscript{2,9}, X.~Defay\textsuperscript{3}, A.~Erb\textsuperscript{3,10}, F.~v.~Feilitzsch\textsuperscript{3}, N.~Ferreiro~Iachellini\textsuperscript{1}, P.~Gorla\textsuperscript{2}, A.~G\"utlein\textsuperscript{4,5}, D.~Hauff\textsuperscript{1}, J.~Jochum\textsuperscript{6}, M.~Kiefer\textsuperscript{1}, H.~Kraus\textsuperscript{7}, J.-C.~Lanfranchi\textsuperscript{3}, A.~Langenk\"amper\textsuperscript{3}, J.~Loebell\textsuperscript{6}, M.~Mancuso\textsuperscript{1}, E.~Mondragon\textsuperscript{3}, A.~M\"unster\textsuperscript{3}, C.~Pagliarone\textsuperscript{2}, F.~Petricca\textsuperscript{1}, W.~Potzel\textsuperscript{3}, F.~Pr\"obst\textsuperscript{1}, R.~Puig\textsuperscript{4,5}, F.~Reindl\textsuperscript{4,5}, J.~Rothe\textsuperscript{1}, K.~Sch\"affner\textsuperscript{2,11}, J.~Schieck\textsuperscript{4,5}, S.~Sch\"onert\textsuperscript{3}, W.~Seidel\textsuperscript{1}\thanks{Deceased}, M.~Stahlberg\textsuperscript{4,5}, L.~Stodolsky\textsuperscript{1}, C.~Strandhagen\textsuperscript{6}, R.~Strauss\textsuperscript{1}, A.~Tanzke\textsuperscript{1}, H.H.~Trinh~Thi\textsuperscript{3}, C. T{\"u}rko{\u{g}}lu\textsuperscript{4,5}, A.~Ulrich\textsuperscript{3}, I.~Usherov\textsuperscript{6}, S.~Wawoczny\textsuperscript{3}, M.~Willers\textsuperscript{3}, M.~W\"ustrich\textsuperscript{1}}\\[1ex] \textsuperscript{1} Max-Planck-Institut f\"ur Physik, D-80805 M\"unchen, Germany \\
\textsuperscript{2} INFN, Laboratori Nazionali del Gran Sasso, I-67010 Assergi, Italy \\
\textsuperscript{3} Physik-Department E15, Technische Universit\"at M\"unchen, D-85747 Garching, Germany \\
\textsuperscript{4} Institut f\"ur Hochenergiephysik der \"OAW, A-1050 Wien, Austria \\
\textsuperscript{5} Atominstitut, Technische Universit\"at Wien, A-1020 Wien, Austria \\
\textsuperscript{6} Eberhard-Karls-Universit\"at T\"ubingen, D-72076 T\"ubingen, Germany \\
\textsuperscript{7} Department of Physics, University of Oxford, Oxford OX1 3RH, United Kingdom \\
\textsuperscript{8} Also at: LIBPhys, Departamento de Fisica, Universidade de Coimbra, P3004 516 Coimbra, Portugal \\
\textsuperscript{9} Also at: Massachusetts Institute of Technology, Cambridge, MA 02139, USA \\
\textsuperscript{10} Also at: Walther-Mei\ss{}ner-Institut f\"ur Tieftemperaturforschung, D-85748 Garching, Germany \\
\textsuperscript{11} Also at: GSSI-Gran Sasso Science Institute, 67100, L'Aquila, Italy\\
}

\contribID{Kluck\_Holger}

\confID{16884}  
\desyproc{DESY-PROC-2017-02}
\acronym{Patras 2017} 
\doi  

\maketitle

\begin{abstract}
CRESST is a multi-stage experiment directly searching for dark matter (DM) using
cryogenic $\mathrm{CaWO_4}$ crystals. Previous stages established leading limits
for the spin-independent DM-nucleon cross section down to DM-particle masses $m_\mathrm{DM}$ 
below $1\,\mathrm{GeV/c^2}$. Furthermore, CRESST performed a dedicated search
for dark photons (DP) which excludes new parameter space
between DP masses $m_\mathrm{DP}$ of $300\,\mathrm{eV/c^2}$ to $700\,\mathrm{eV/c^2}$.

In this contribution we will discuss the latest results based on the previous
CRESST-II phase 2 and we will report on the status of the current CRESST-III
phase 1: in this stage we have been operating 10 upgraded detectors with
$24,\mathrm{g}$ target mass each and enhanced detector performance since summer 2016. The improved detector
design in terms of background suppression and reduction of the detection
threshold will be discussed with respect to the previous stage.
We will conclude with an outlook on the potential of the next stage, CRESST-III
phase 2.
\end{abstract}

\section{Introduction}
The nature of dark matter is one of the greatest mysteries in modern physics:
albeit observations ranging from galactic dynamics to the cosmic microwave
background indicate evidently that most matter is dark \cite{bertone2010}, no
particle constituent for dark matter was unambiguously found.
Whereas few experiments claim a potential signal, e.g.\ DAMA/LIBRA
\cite{dama}, the majority observed null signals, e.g. LUX \cite{lux}, SuperCDMS
\cite{scdms}, and CRESST \cite{tum40,lise}.
Besides the classic WIMP, also lighter candidates were discussed during the
last years, e.g.\ \textit{asymmetric dark matter} \cite{adm} with masses
$m_\mathrm{DM}$ in the order of $\mathcal{O}(\mathrm{GeV/c^2})$ or \textit{dark
photons} \cite{darkphoton1,darkphoton2} with masses $m_\mathrm{DP}$ on the
keV-scale and below.
Scattering of the former with target nuclei may cause
nuclear recoils, whereas the latter may interact electromagnetically with
electrons.

In this contribution we will show that CRESST is ideally suited to search
for light dark matter particles. After a short introduction of the CRESST
experiment in Sec.~\ref{sec:cresst}, we report the latest results based on
CRESST-II phase~2 data, both for the search for dark photons
(Sec.~\ref{sec:darkphotons}) and for the search of dark matter induced
nuclear recoils (Sec.~\ref{sec:cresst2}).
Afterwards, we discuss the current status and
potential of CRESST-III in Sec.~\ref{sec:cresst3}.

\section{\label{sec:cresst}The CRESST experiment}
The multi-stage \textit{Cryogenic Rare Event Search with Superconducting
Thermometers} (CRESST), based at the \textit{Laboratori Nazionali del Gran Sasso} (LNGS) underground laboratory in Italy, mainly looks for
nuclear recoils induced by elastic scattering of dark matter particles in its
target: scintillating $\mathrm{CaWO_4}$ crystals which are operated
at temperatures of the order of $\mathcal{O}(10\,\mathrm{mK})$. By 
simultaneously reading out two signal channels, CRESST is able to distinguish
nuclear recoils from electromagnetic interactions.
Interactions in the target create non-thermal phonon excitations of the
crystal lattice (\textit{phonon signal}) which are recorded by a
\textit{transition edge sensor} (TES) as well as scintillation light
(\textit{light signal}) recorded by a separate light detector. The scintillation
efficiency depends on the type of interaction:
for nuclear recoils it is quenched with respect to
electromagnetic interactions, i.e.\ for a given phonon signal the light signal is reduced. A
cut on the light-to-phonon signal ratio efficiently selects nuclear 
recoils against electromagnetic interactions, since the energies of all types of interactions are precisely
measured by the phonon signal.
A detailed description can be found in \cite{tum40,lise} and references therein. 

\section{Latest results from CRESST-II phase 2}
CRESST-II phase 2 operated 18 target crystals, each
encapsulated together with a light detector in a detector module. Here, we
focus on results obtained by two of these modules, \textit{TUM40} \cite{tum40}
and \textit{Lise} \cite{lise}.

\subsection{\label{sec:darkphotons}Search for dark photons}
\begin{wrapfigure}[22]{R}{0.55\textwidth}
\centerline{
\includegraphics[width=0.54\textwidth]{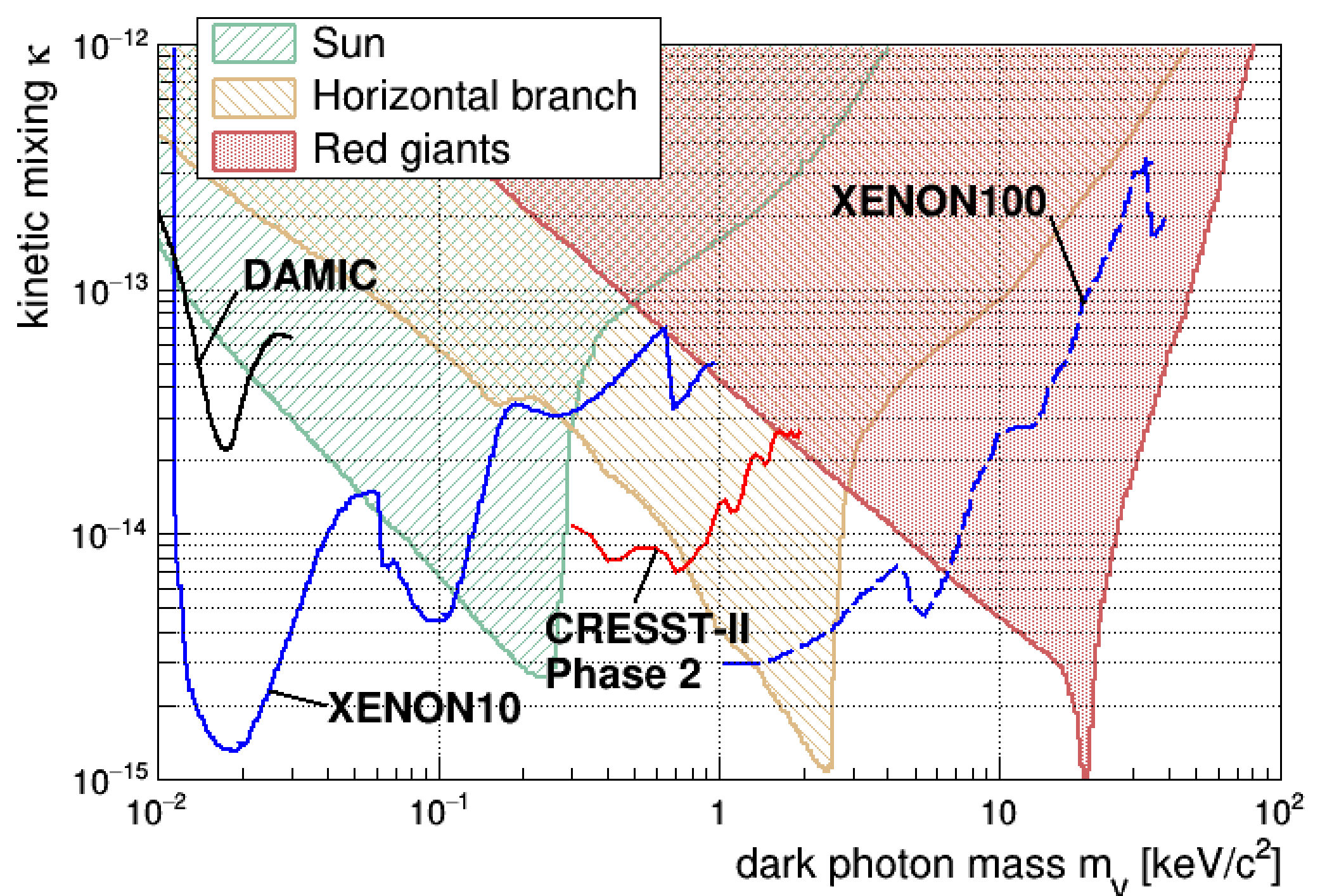}
}
\caption{\label{cresstphotons:exclusionplot}Parameter space for the kinetic
mixing between dark photons and standard model photons.
Shown in \textit{red} are the exclusion limit for CRESST-II phase~2 obtained
with the detector module Lise in comparison with results from LXe and Si
based experiments together with constraints from anomalous energy
loss in stars.
For references see \protect\cite{cresstphoton}.}
\end{wrapfigure}
Dark photons are long-lived vector particles which may constitute dark matter as
discussed in \cite{darkphoton1,darkphoton2}. Via kinetic mixing $\kappa$ with
standard model photons they may get absorbed with a cross-section
$\sigma_\mathrm{DP}\approx\kappa c^2 \sigma_\gamma$ approximately proportional
to the photoelectric cross-section $\sigma_\gamma$. Due to the negligible velocity of
galactic dark matter the resulting signature would be a distinct peak at the
rest mass $m_\mathrm{DP}c^2$ of the dark photon.

For our dark photon search \cite{cresstphoton} we used the same data set of
detector module Lise which we also used for a search for dark matter-nucleon
scatterings in \cite{lise}. Contrary to the search for nuclear recoils in
\cite{lise}, the light-to-phonon signal ratio was inefficient to reject
backgrounds because both, a potential dark photon signal and radioactive decays
as the main background component, result in recoiling electrons.

To search for a potential signal peak at a given rest mass
$m_\mathrm{DP}c^2$, a Gaussian together with an
empirical background model was fitted to the data using a Bayesian approach. Repeating
this procedure in $50\,\mathrm{eV}$ steps between $0.3\,\mathrm{keV}$ and
$2\,\mathrm{keV}$ resulted in an improved limit on the kinetic mixing $\kappa$
\cite{cresstphoton} as shown in Fig.~\ref{cresstphotons:exclusionplot}: between $300$ and
$700\,\mathrm{eV}/c^2$ CRESST could exclude new parameter
space.

\subsection{\label{sec:cresst2}Search for dark matter-nucleon scatterings}

\begin{wrapfigure}[22]{r}{0.55\textwidth}
\centerline{
\includegraphics[width=0.54\textwidth]{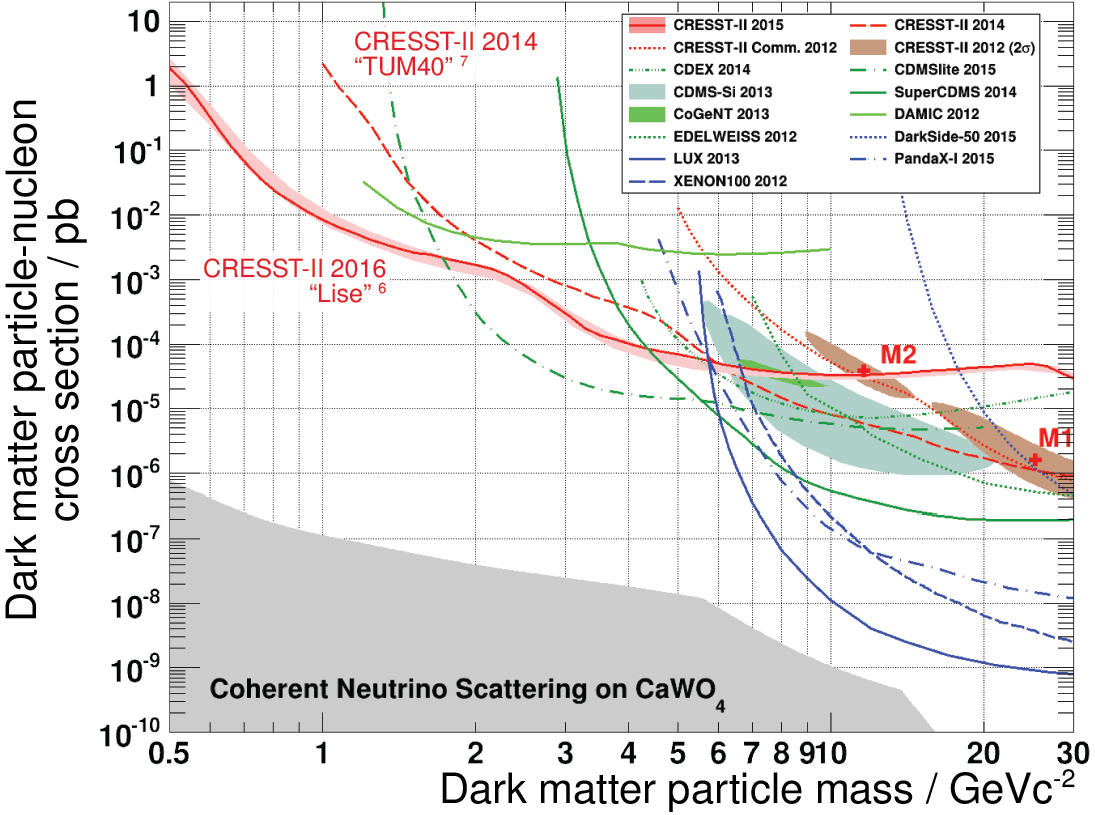}
}
\caption{\label{cresst:exclusionplot}Parameter space for elastic
spin-independent dark matter-nucleon scattering. Shown in \textit{red} are the
exclusion limits for CRESST-II phase~2 obtained with the detector modules
TUM40 and Lise in comparison with results from LAr, LXe, Ge, and Si
based experiments. For references see \protect\cite{lise}.}
\end{wrapfigure}
At the respective time of
publication, CRESST-II phase~2 set leading
limits on the spin-independent cross section for elastic dark matter-nucleon
scattering with the detector modules \textit{TUM40} and \textit{Lise}, the latter reaching down to $m_\mathrm{DM} =
500\,\mathrm{MeV/c^2}$, see Fig.~\ref{cresst:exclusionplot}.
Both detector modules exemplify different approaches to increase the detection
sensitivity for dark matter induced nuclear recoils: reduction of the background
and lowering of the detection threshold.

With TUM40 we could reduce the background rate due to intrinsic
contamination of the $\mathrm{CaWO_4}$ down to ${\sim 1}\mhyphen
3\,\mathrm{mBq/kg}$ \cite{tum40}.
The additional background due to surface-alpha events was
completely rejected by an active veto based on scintillating
$\mathrm{CaWO_4}$ sticks to hold the target \cite{stick}.

Because the nuclear recoil spectrum gets steeper with decreasing dark
matter-particle mass, a reduced detection threshold increases especially
the sensitivity for low-mass dark matter candidates. This is highlighted by the
detector module Lise which featured  with $307\,\mathrm{eV}$ the lowest
detection threshold within CRESST-II phase~2 \cite{lise}, cf.\ Fig.~\ref{cresst:exclusionplot}.

\section{\label{sec:cresst3}Status and potential of CRESST-III}
To further increase its sensitivity for low-mass dark matter, CRESST in
its third stage aims at a combination of both approaches: detector modules with
high radiopurity and low detection threshold. CRESST-III phase 1 is expected to reach a detection
threshold of $100\,\mathrm{eV}$. In phase 2 we plan to reduce the background
by a factor of 100 compared to TUM40 \cite{cresst3}.

CRESST-III phase~1 is equipped with 10 CRESST-III detector modules, each
featuring $24\,\mathrm{g}$ of target mass of at least TUM40 quality and each held by
iSticks: an active rejection technique against holder
related backgrounds e.g.\ relaxation events \cite{prototype}.
Already 2 targets are produced from $\mathrm{CaWO_4}$ crystals which were grown
from chemically purified raw materials based on the ongoing
R\&D for CRESST-III phase~2. As prototype measurements \cite{prototype} showed,
the decreased target mass will lower the detection threshold towards
$100\,\mathrm{eV}$ via a reduced heat capacity.

Since August 2016 CRESST is cooled down to operational temperature and records
data. After an extensive gamma-calibration campaign in October 2016, we take
physics data since November 2016, interrupted by a neutron calibration campaign
in April 2017.
It is planned to continue data taking for one year with all 10 modules, aiming
for an total exposure of $50\,\mathrm{kg \cdot d}$. 80\% of the data are dynamically
blinded, allowing the preparation of selection cuts on the remaining 20\% which
will be excluded from the final analysis.

Keeping in mind that CRESST-III aims for unprecedented low detection thresholds
and under the premise that no so far unknown background appears at these energies
 we expect to reach a sensitivity of ${\sim}
5\cdot 10^{-5}\,\mathrm{pb}$ at $1\,\mathrm{GeV/c^2}$ with CRESST-III phase~1 \cite{cresst3}.
This would be an improvement by roughly 2 orders of magnitude compared to the
Lise result of \mbox{CRESST-II} phase~2.
In CRESST-III phase~2 we plan to operate 100 detector modules for 2 years,
in order to reach an exposure of $1000\,\mathrm{kg \cdot d}$. With this exposure, CRESST will
be close to the \textit{neutrino floor} of $\mathrm{CaWO_4}$ \cite{cresst3} where the coherent
scattering of solar neutrinos on $\mathrm{CaWO_4}$ will become a significant
background for any future search for low-mass dark matter.


\begin{footnotesize}

\end{footnotesize}


\end{document}